\documentclass[12pt,doublecol]{article}
\usepackage[cp1251]{inputenc}
\usepackage{amsmath,amssymb,amsfonts,graphicx,indentfirst}
\usepackage[a4paper]{geometry}
\usepackage{caption2}
\usepackage {mathrsfs}
\usepackage {cite}

\begin{document}
\begin{center}{\large\bf The influence of Coulomb correlations on nonequilibrium quantum transport in quadruple quantum-dot structure}\end{center}
\begin{center}{M.Yu. Kagan$^{1,2}$, S.V. Aksenov$^{3}$}\end{center}
\begin{center}{kagan@kapitza.ras.ru, asv86@iph.krasn.ru}\end{center}
\begin{center}{\small$^1$P.L. Kapitza Institute for Physical Problems RAS - 119334 Moscow, Russia\\
\small$^2$National Research University Higher School of Economics - 101000 Moscow, Russia\\
\small$^3$Kirensky Institute of Physics, Federal Research Center KSC SB RAS - 660036 Krasnoyarsk, Russia}\end{center}

PACS: {$73.21$.La - Quantum dots,\\
$73.63$.Kv - Electronic transport in QDs,\\
$85.35$.-p - Nanoelectronic devices}

\abstract{The description of quantum transport in a quadruple quantum-dot structure (QQD) is proposed taking into account the Coulomb correlations and nonzero bias voltages. To achieve this goal the combination of nonequilibrium Green's functions and equation-of-motion technique is used. It is shown that the anisotropy of kinetic processes in the QQD leads to negative differential conductance (NDC). The reason of the effect is an interplay of the Fano resonances which are induced by the interdot Coulomb correlations. Different ways to increase the peak-to-valley ratio related to the observed NDC are discussed.}


\section{Introduction}

Technological development in recent decades allowed the experimental study of systems of few-electron quantum dots \cite{elzerman-03,gaudreau-06}. In these structures the occupation of each dot and the interaction between them are governed by the electric fields of gate electrodes. Since the lifetime of single-electron spin state, a spin qubit, in semiconducting quantum dot is relatively long, such objects are attractive for storage and processing of quantum information \cite{loss-98,koppens-06}. The research of quantum-dot complexes in this direction is necessary to create a scalable architecture of spin qubits \cite{delbecq-14,ito-16}.

Intra- and interdot Coulomb interactions are the key factors determining different many-particle effects in the systems of quantum dots. They are being considered as a perfect testbed to study the properties of the Hubbard model due to the possibility of effective control of the internal parameters such as the interdot tunneling, single-electron energies and intensities of Coulomb interactions \cite{byrnes-08}.

Nowadays the structures consisting of three and four quantum dots are experimentally available and can be studied in different topologies. The dots can form a linear molecule where the nearest-neighbor tunneling of electrons takes place and the edge dots are disconnected from each other. Alternatively, such dots can be arranged in the shape of triangle or square, respectively. In this case there is the nearest-neighbor coupling of all the dots \cite{hsieh-12a,thalineau-12,baart-16}. The topology significantly affects the system properties. In particular, considering the Hubbard model with very large values of the intradot Coulomb repulsion, $U$, it was shown that the presence of closed paths for the motion of electrons allows realization of Nagaoka ferromagnetic order \cite{oguri-07,barthelemy-13}. In case of quadruple quantum-dot structure (QQD) with three electrons the appearance of ground state with spin $S=3/2$ is explained by the presence of effective gauge field which leads to an increase of the energy of chiral state with spin $S=1/2$. This effect is one of the mechanisms that initiates a spin blockade of electron current through the QQD \cite{ozfidan-13}. In this regime the transitions between the states which differ in the number of electrons by $1$ are forbidden if the spin of these states differs by more than $1/2$. It should be noted that the spin blockade was also demonstrated earlier for double- and triple linearly connected quantum dots \cite{johnson-05,hsieh-12b} and for a separate multi-level dot \cite{weinmann-95}. One of its manifestations in the observable values is a current rectification and a negative differential conductance (NDC). Among other mechanisms of current suppression in quantum-dot systems one can mention the Aharonov-Bohm effect \cite{urban-09}, the dark states \cite{michaelis-06,poltl-09,maslova-16} and the isospin blockade \cite{jacob-04}.

\begin{figure}\center
\includegraphics[width=0.6\textwidth]{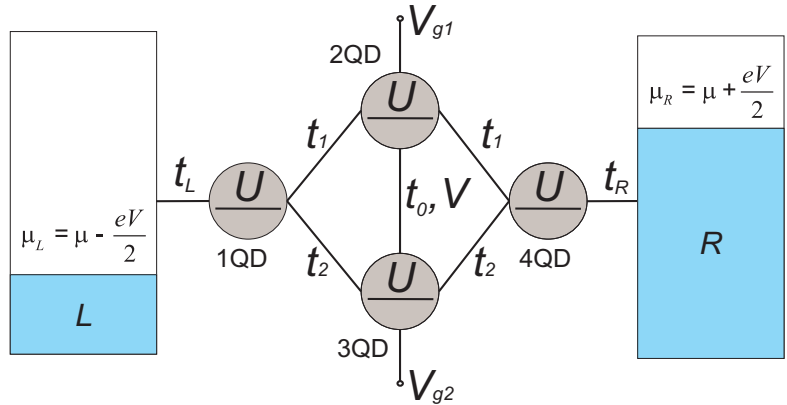}
\caption{The QQD between the one-band paramagnetic leads.}
\label{fig.1}
\end{figure}
In this article we propose an alternative description of the NDC effect observed in the transport properties of QQD. The investigated system is schematically shown in Figure \ref{fig.1}. The dots constituting the device are located at the vertices of square. The left and right metal contacts considered in the one-band approximation are connected to the 1st and 4th dots (1QD and 4QD), respectively. Thus, there are second and third dots (2QD and 3QD) in the central part of the QQD and two paths, top and bottom ones, for electron transport. The electron current is found as a result of solving the systems of equations of motion for the nonequilibrium Green's functions. The NDC effect arising in the case of the anisotropic QQD is interpreted in terms of bound states in continuum (BICs) and the interaction of Fano resonances which are formed by the Coulomb correlations between the electrons of central dots, $V$.

\section{Model}

The Hamiltonian of QQD between the metallic leads is $\hat{H}=\hat{H}_L+\hat{H}_R+\hat{H}_D+\hat{H}_T$. The terms $\hat{H}_L$ and $\hat{H}_R$ describe left and right one-band leads, $\hat{H}_{\alpha}=\sum_{k\sigma}\xi_{k\sigma}c^{+}_{\alpha k\sigma}c_{\alpha k\sigma}$, where $c_{\alpha k\sigma}$ annihilates an electron with a wave vector $k$, spin projection $\sigma$ and energy $\xi_{k\sigma}=\epsilon_{k\sigma}-\mu$ ($\mu=0$ is a chemical potential) in the $\alpha$-th lead ($\alpha=L,R$).

The QQD Hamiltonian reads
\begin{eqnarray} \label{H4l}
&&\hat{H}_{QQD} =\sum\limits_{\sigma;j=1}^{4}\xi_{j\sigma}a^+_{j\sigma}a_{j\sigma}+U\sum\limits_{j=1}^{4}n_{j\uparrow}n_{j\downarrow}\nonumber\\
&&+V\sum\limits_{\sigma\sigma'}n_{2\sigma}n_{3\sigma'}+ \sum\limits_{\sigma}\left[t_{1}\left(a^+_{1\sigma}+a^+_{4\sigma}\right)a_{2\sigma}\right.\\
&&\left.+t_{2}\left(a^+_{1\sigma}+a^+_{4\sigma}\right)a_{3\sigma}+t_{0}a^{+}_{2\sigma}a_{3\sigma}+h.c.\right],~\nonumber
\end{eqnarray}
where $a_{j\sigma}$ annihilates an electron with a spin projection $\sigma$ and an energy $\xi_{j\sigma}=\epsilon_{j\sigma}-\mu$ on the level of $j$-th dot; $t_{1,2}$ - a hopping parameter in the top (1QD-2QD-4QD) or bottom (1QD-3QD-4QD) arms (see fig. \ref{fig.1}); $t_0$ - a hopping parameter between the arms; $U,~V$ - intensities of intra- and interdot Coulomb repulsion, respectively.

The last term in the Hamiltonian $\hat{H}$ is responsible for the interaction between the subsystems,
\begin{equation} \label{HT1}
\hat{H}_T =T_{L}\left(t\right)\sum\limits_{k\sigma}c_{Lk\sigma}^{+}a_{1\sigma} +
T_{R}\left(t\right)\sum\limits_{k\sigma}c_{Rk\sigma}^{+}a_{4\sigma} + h.c.,
\end{equation}
where $T_{L\left(R\right)}\left(t\right)=t_{L\left(R\right)}e^{\mp\frac{ieV}{2}t}$ - a coupling parameter of the QQD with the left (right) lead. Note that the time dependence in $T_{L\left(R\right)}\left(t\right)$ appears due to nonequilibrium conditions meaning that the electrochemical potentials, $\mu_L$ and $\mu_R$, are different from each other, $\mu_R-\mu_L=eV$ \cite{rogovin-74}. In subsequent calculations of the current and conductance we consider symmetric transport regime, $t_L=t_R=t$.

\section{Electric current}

The operator of steady-state electric current is defined as $\left\langle I\left(t,t\right)\right\rangle \equiv I=e\left\langle \dot{N}_L\right\rangle$, where $N_L=\sum_{k\sigma}c^{+}_{Lk\sigma}c_{Lk\sigma}$ - the left-lead particle operator. Writing the equation of motion one can get ($\hbar=1$)
\begin{equation} \label{IL1}
I=ie\sum_{k\sigma}\Biggl[ T_{L}^{+}\left(t\right)G_{Lk1\sigma}^{+-}\left(t,t\right)
-T_{L}\left(t\right)G_{1Lk\sigma}^{+-}\left(t,t\right)\Biggr].
\end{equation}
The nonequilibrium Green's functions are introduced in the expression (\ref{IL1}). The operators $\psi_{n}=c_{\alpha k\sigma},a_{j\sigma}$ entering into them are ordered on the Keldysh contour, $C$ \cite{keldysh-65}.

If (\ref{HT1}) is treated as an interaction operator than the analysis of perturbation-theory series for the functions $G_{Lk1\sigma}^{+-}$ and $G_{1Lk\sigma}^{+-}$ results in the following formula for the current,
\begin{eqnarray} \label{IL2}
&&I=e\sum_{\sigma}\int\limits_{C}d\tau_{1}\Bigl[\Sigma_{L\sigma}^{+a}\left(t-\tau_{1}\right)G_{11\sigma}^{a-}\left(\tau_{1}-t\right)-\\
&&~~~~~~~~~~~~~~~~~~~~~~~~~-G_{11\sigma}^{+a}\left(t-\tau_{1}\right)\Sigma_{L\sigma}^{a-}\left(\tau_{1}-t\right)\Bigr],~\nonumber
\end{eqnarray}
where the self-energy functions are introduced, $\Sigma_{L\sigma}^{ab}\left(\tau-\tau'\right)=T_{L}^{+}\left(\tau\right)
\sum_{k}g_{Lk\sigma}^{ab}\left(\tau-\tau'\right)T_{L}\left(\tau'\right)$, which characterize the influence of the left lead on the QQD; $g_{Lk\sigma}^{ab}\left(\tau-\tau'\right)$ - the one-electron Green's function of the left lead. The value of upper indexes, $a,b=+,-$, points out the branch of Keldysh contour, $C_{+}$, $C_{-}$. The general form of the Dyson equation for the Green's function $G_{11\sigma}\left(\tau-\tau'\right)$ is
\begin{eqnarray} \label{Dys11}
&&G_{11\sigma}\left(\tau-\tau'\right)=g_{11\sigma}\left(\tau-\tau'\right)+\\
&&+\int\limits_{C}d\tau_{1}\tau_{2}\Bigl[g_{11\sigma}\left(\tau-\tau_{1}\right)\Sigma_{L\sigma}\left(\tau_{1}-\tau_{2}\right)G_{11\sigma}\left(\tau_{2}-\tau'\right)+\Bigr.\nonumber\\
&&\Bigl.~~~~~~~~~~~~~~+g_{14\sigma}\left(\tau-\tau_{1}\right)\Sigma_{R\sigma}\left(\tau_{1}-\tau_{2}\right)G_{41\sigma}\left(\tau_{2}-\tau'\right)\Bigr],~\nonumber
\end{eqnarray}
where $g_{ij\sigma}\left(\tau-\tau'\right)$ - the bare Green's functions of the QQD. During the derivation of (\ref{IL2}) and (\ref{Dys11}) we deal with the nonmagnetic approximation. Specifically, the spin-flip processes are neglected, $\langle a_{i\sigma} a_{j\overline{\sigma}}^{+} \rangle=0$. After the transition to integration over the real time contour and the subsequent Fourier transform we obtain the following expression
\begin{equation} \label{IL3}
I=i\Gamma\sum_{\sigma}\int\limits_{-\infty}^{+\infty}d\omega\Bigl[f_L\left(G_{11\sigma}^{a}-G_{11\sigma}^{r}\right)-G_{11\sigma}^{+-}\Bigr],
\end{equation}
where $f_L\equiv f\left(\omega+\frac{eV}{2}\right)$ - the Fermi-Dirac distribution function; $\Gamma/2=\Gamma_{L}=\Gamma_{R}=\pi t^2 g$ - the parameter that describes the broadening of QQD levels due to the coupling with the leads. In general, the density of states of lead depends on frequency and spin projection, $g_{\sigma}\left(\omega\right)=\sum_{k}\delta\left(\omega-\xi_{k\sigma}\right)$. However, in the article the leads are supposed to be paramagnetic and have wide band. Consequently, these dependencies can be ignored and $g=const$. As a result, the Fourier transforms of self-energy functions of $\alpha$-th lead are $\Sigma_{\alpha\sigma}^{r}=-\frac{i}{2}\Gamma$ and $\Sigma_{\alpha\sigma}^{+-}=i\Gamma f_{\alpha}$.

To obtain the final expression describing the steady-state current in the system let us find the Green's functions of the QQD entering into (\ref{IL3}). For this purpose we use the equation-of-motion technique. The general form of equations for $G_{i\sigma j\sigma'}^{r}\left(\omega\right)\equiv\langle\langle a_{i\sigma} | a_{j\sigma'}^{+} \rangle\rangle^{r}$ and $G_{i\sigma j\sigma'}^{+-}\left(\omega\right)\equiv\langle\langle a_{i\sigma} | a_{j\sigma'}^{+} \rangle\rangle^{+-}$ differs from each other because of the definition of $G_{i\sigma,j\sigma'}^{r,+-}\left(t-t'\right)$,
\begin{eqnarray} \label{eqG1}
&&z\langle\langle a_{i\sigma} | a_{j\sigma'}^{+} \rangle\rangle^{r}=\left\langle\left\{a_{i\sigma},~a_{j\sigma'}^{+}\right\}\right\rangle+
\langle\langle \left[a_{i\sigma},~\hat{H}\right] | a_{j\sigma'}^{+} \rangle\rangle^{r},\nonumber\\
&&z\langle\langle a_{i\sigma} | a_{j\sigma'}^{+} \rangle\rangle^{+-}=
\langle\langle \left[a_{i\sigma},~\hat{H}\right] | a_{j\sigma'}^{+} \rangle\rangle^{+-},\nonumber
\end{eqnarray}
where $z=\omega+i\delta$. In addition, taking into account the diagram expansion of mixed Green's function, $G_{L\left(R\right)kj\sigma}\left(t-t'\right)=\int\limits_{C}g_{L\left(R\right)kj\sigma}
\left(t-\tau\right)T_{L\left(R\right)}\left(\tau\right)G_{1\left(4\right)j\sigma}
\left(\tau-t'\right)$, the corresponding equations become
\begin{eqnarray} \label{eqGLRD}
z\langle\langle c_{L\left(R\right)k\sigma} | a_{j\sigma}^{+} \rangle\rangle^{r}=g_{L\left(R\right)k\sigma}^{r}t_{L\left(R\right)}
\langle\langle a_{1\left(4\right)\sigma} | a_{j\sigma}^{+} \rangle\rangle^{r},
~~~~~~~~~~~\nonumber\\
z\langle\langle c_{L\left(R\right)k\sigma} | a_{j\sigma'}^{+} \rangle\rangle^{+-}=
~~~~~~~~~~~~~~~~~~~~~~~~~~~~~~~~~~~~~~~~~~~~\nonumber\\
t_{L\left(R\right)}
\left(g_{L\left(R\right)k\sigma}^{r}
\langle\langle a_{1\left(4\right)\sigma} | a_{j\sigma}^{+} \rangle\rangle^{+-}+g_{L\left(R\right)k\sigma}^{+-}
\langle\langle a_{1\left(4\right)\sigma} | a_{j\sigma}^{+} \rangle\rangle^{a}\right),\nonumber
\end{eqnarray}
where $g_{\alpha k\sigma}^{r}=\left(z-\xi_{k\sigma}\right)^{-1}$, $g_{\alpha k\sigma}^{+-}=2\pi i f_{\alpha}\delta\left(\omega-\xi_{k\sigma}\right)$. Next, to derive the closed systems of equations we use the decoupling procedure for the nonmagnetic case developed in the refs. \cite{you-99a,kagan-17a,kagan-17b}. Such an approximation is valid at temperatures higher than the Kondo temperature \cite{lacroix-81}. In the employed approach the equations for the third-order Green's functions, e.g. $\langle\langle n_{3\sigma}n_{2\overline{\sigma}}a_{2\sigma} | a_{j\sigma}^{+} \rangle\rangle^{r,+-}$, should be decoupled. The solution of the final set of equations for the retarded Green's functions is \begin{eqnarray} \label{Gr}
G_{\beta\beta}^r=\frac{C_{\beta}Z_{\overline{\beta}}}{Z},~G_{\beta\overline{\beta}}^r=
\frac{C_{\beta}C_{\overline{\beta}}x_{2}}{Z},~G_{\alpha\alpha}^r=
\frac{C_{\alpha}\Delta_{\overline{\alpha}}}{Z},
~~~~~~~~~~\\
G_{\alpha\overline{\alpha}}^r=\frac{C_{\alpha}C_{\overline{\alpha}}\Delta_{1}}{Z},
G_{\beta\alpha}^r=\frac{C_{\alpha}C_{\beta}T_{\overline{\beta}}P_{\overline{\alpha}}}{Z},
\beta\left(\alpha\right)=1,4\left(2,3\right),\nonumber
\end{eqnarray}
where $\Delta_{\alpha}=D_{\alpha}T_{\beta}T_{\overline{\beta}}-t^{2}\left(\alpha\right)C_{\alpha}S$, $\Delta_{1}=t_{0}T_{\beta}T_{\overline{\beta}}+t\left(\alpha\right)t\left(\overline{\alpha}\right)S$, $S=C_{\beta}T_{\overline{\beta}}+C_{\overline{\beta}}T_{\beta}$, $P_{\alpha}=t\left(\overline{\alpha}\right)D_{\alpha}+t_{0}t\left(\alpha\right)C_{\alpha}$, $Z=T_{\beta}T_{\overline{\beta}}x_{1}-Sx_{2}$, $Z_{\beta}=T_{\beta}x_{1}-C_{\beta}x_{2}$, $T_{\beta}=D_{\beta}+i\Gamma C_{\beta}/2$, $x_{1}=\Delta_{\alpha}\Delta_{\overline{\alpha}}-t_{0}^{2}C_{\alpha}C_{\overline{\alpha}}$, $x_{2}=t\left(\alpha\right)C_{\alpha}P_{\overline{\alpha}}+t\left(\overline{\alpha}\right)C_{\overline{\alpha}}P_{\alpha}$, $t\left(\alpha\right)=t_{1,2}$. The factors $C_{\alpha,\beta}$ and $D_{\alpha,\beta}$ contain the explicit dependencies on the occupation numbers, correlators and intensities of the Coulomb interactions in the QQD: $C_{\alpha}=C_{\alpha1}+C_{\alpha2}$, $C_{\alpha1}=b_{\alpha4}\left(b_{\alpha2}b_{\alpha3}+Ub_{\alpha3}\langle n_{\alpha}\rangle+2Vb_{\alpha2}\langle n_{\overline{\alpha}}\rangle\right)$, $C_{\alpha2}=UV\left(b_{\alpha2}+b_{\alpha3}\right)\left(2\langle n_{\alpha}\rangle\langle n_{\overline{\alpha}}\rangle-\langle a_{\alpha}^{+}a_{\overline{\alpha}}\rangle^{2}\right)$, $C_{\beta}=b_{\beta2}+U\langle n_{\beta}\rangle$,  $D_{\alpha}=b_{\alpha1}b_{\alpha2}b_{\alpha3}b_{\alpha4}$, $D_{\beta}=b_{\beta1}b_{\beta2}$, $b_{\alpha1}=z-\xi_{\alpha}$, $b_{\alpha2}=b_{\alpha1}-U$, $b_{\alpha3}=b_{\alpha1}-V\left(1+\langle n_{\overline{\alpha}}\rangle\right)$, $b_{\alpha4}=b_{\alpha3}-U$. Note that in the formulas (\ref{Gr}) for simplicity the spin indexes are omitted as in the nonmagnetic case we have $\langle a_{i\sigma}^{+}a_{j\sigma}\rangle=\langle a_{i\overline{\sigma}}^{+}a_{j\overline{\sigma}}\rangle$. In turn, the solution of the system of equations for $G^{+-}_{ij}$ gives
\begin{eqnarray} \label{Gpm}
&&G_{\beta\beta\left(\overline{\beta}\right)}^{+-}=i\Gamma\frac{C_{\beta}\left(f_{\beta}Z_{\overline{\beta}}G_{\beta\beta\left(\overline{\beta}\right)}^{a}+f_{\overline{\beta}}C_{\overline{\beta}}x_{2}G_{\overline{\beta}\beta\left(\overline{\beta}\right)}^{a}\right)}{Z},\nonumber\\
	 &&G_{\alpha\alpha\left(\overline{\alpha}\right)}^{+-}=i\Gamma\frac{C_{\alpha}P_{\overline{\alpha}}\left(f_{\beta}C_{\beta}T_{\overline{\beta}}G_{\beta\alpha\left(\overline{\alpha}\right)}^{a}+f_{\overline{\beta}}C_{\overline{\beta}}T_{\beta}G_{\overline{\beta}\alpha\left(\overline{\alpha}\right)}^{a}\right)}{Z},\nonumber\\
	 &&G_{\beta\alpha}^{+-}=i\Gamma\frac{C_{\beta}\left(f_{\beta}Z_{\overline{\beta}}G_{\beta\alpha\left(\overline{\beta}\right)}^{a}+f_{\overline{\beta}}C_{\overline{\beta}}x_{2}G_{\overline{\beta}\alpha\left(\overline{\beta}\right)}^{a}\right)}{Z},
\end{eqnarray}
where  $f_{\beta}=f_{L,R}$, $G_{ij}^{a}=\left(G_{ij}^{r}\right)^{*}$, $G_{ij}^{+-}=-\left(G_{ji}^{+-}\right)^{*}$. Proceeding from the definition of lesser Green's functions the correlators and occupation numbers can be obtained by self-consistent solution of the following integral equations
\begin{equation} \label{kineq}
\langle n_{i}\rangle=2\int\limits_{-\infty}^{+\infty}\frac{d\omega}{2\pi}G_{ii}^{+-},~\langle a_{i}^{+}a_{j}\rangle=2\int\limits_{-\infty}^{+\infty}\frac{d\omega}{2\pi}G_{ji}^{+-}.
\end{equation}
Substituting the calculated Green's functions in (\ref{IL3}) we find the final expression describing the current in the QQD,
\begin{eqnarray} \label{IL4}
&&I=2e\Gamma^2\int\limits_{-\infty}^{+\infty}d\omega G_{14}^{r}G_{41}^{a}\left(f_{L}-f_{R}\right)=\\
&&~~~~~~~~~~~~~~~~~~=2e\Gamma^2\int\limits_{-\infty}^{+\infty}d\omega \frac{C_{1}^{2}C_{4}^{2}x_{2}^{2}}{|Z|^2}\left(f_{L}-f_{R}\right).\nonumber
\end{eqnarray}
Note that the factor $2$ in the numerators of formulas (\ref{kineq}) and (\ref{IL4}) arises as a result of the summation over the spin indexes. In further discussion all the energy values are measured in units of $\Gamma$. Additionally, the regime of strong coupling with contacts will be analyzed ($\Gamma=t_1$). In subsequent calculations one-electron energies of the edge dots are assumed to be the same, $\xi_{1\sigma}=\xi_{4\sigma}=\varepsilon_D$. The difference of energies of two central dots is controlled by the parameter $\Delta$, $\xi_{2\left(3\right)\sigma}=\varepsilon_D\pm\Delta$.

\begin{figure}\center
\includegraphics[width=0.45\textwidth]{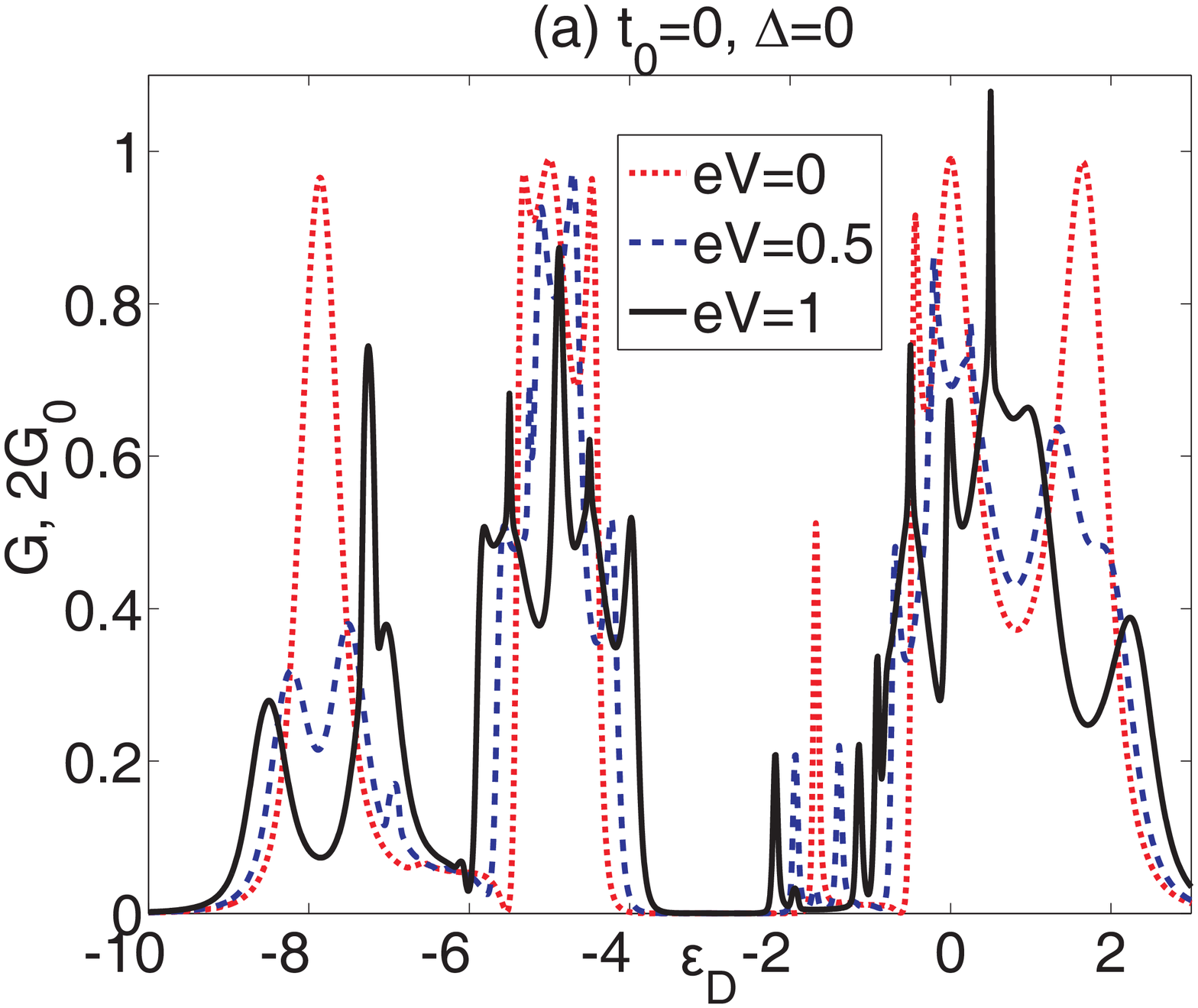}
\includegraphics[width=0.45\textwidth]{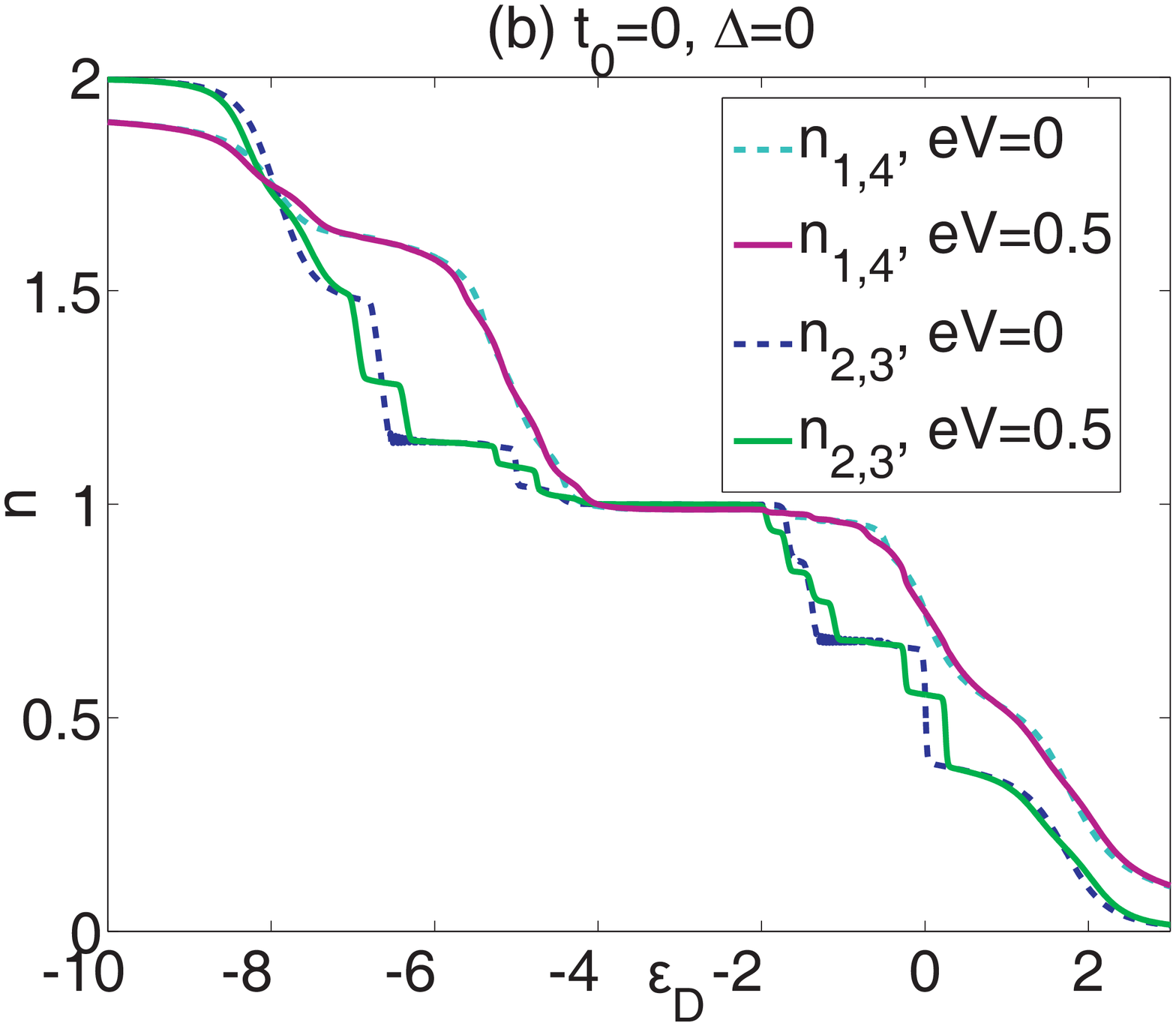}
\caption{The effect of bias voltage on the gate-voltage dependencies of conductance (a) and occupation numbers (b) for the isotropic QQD. Parameters: $U=5$, $V=1$, $t_1=t_2=1$, $t_0=\Delta=0$, $k_{B}T=0.01$.}\label{fig.2}
\end{figure}

\section{Results and discussion}

We now turn to the description of nonequilibrium transport through the QQD. The figures \ref{fig.2}a and b show the conductance of QQD and its occupation numbers as functions of gate field at different bias voltages for the isotropic case, $t_1=t_2$. It is seen that the resonances of $G\left(\varepsilon_D\right)$, which are located left and right from the insulating band (that corresponds to the half-filling), are split in comparison with the equilibrium regime (compare, e.g dotted and dashed curves in Fig. \ref{fig.2}a). It can be explained by the fact that for $eV\neq0$ the transmission of electrons is enhanced if the QQD energy level governed by the parameter $\varepsilon_D$ coincides with the electrochemical potential of the left or right lead, $\mu_{L,R}=\mu\mp\frac{eV}{2}$. Simultaneously, the Fano antiresonances in the conductance emerging due to the Coulomb interaction between the central dots \cite{kagan-17a,kagan-17b} are modified if $eV\neq0$. Both insulating bands obtained in the linear response regime persist at $eV=0.5$. However, the further increase in bias voltage gives rise to the decrease of the bands' widths (solid curve in Fig. \ref{fig.2}a). Moreover, in strongly nonequilibrium regime the effects which cannot be described by the Landauer-Buttiker formula may appear. As a result, for some gate voltages in the situation when $\Gamma \sim U,~V$: $G>2G_0$. The steps of occupation numbers are also split at $eV\neq0$ which is especially evident for the populations of two internal dots (Fig. \ref{fig.2}b). In this case each step corresponds to the conductance resonance.

\begin{figure}\center
\includegraphics[width=0.45\textwidth]{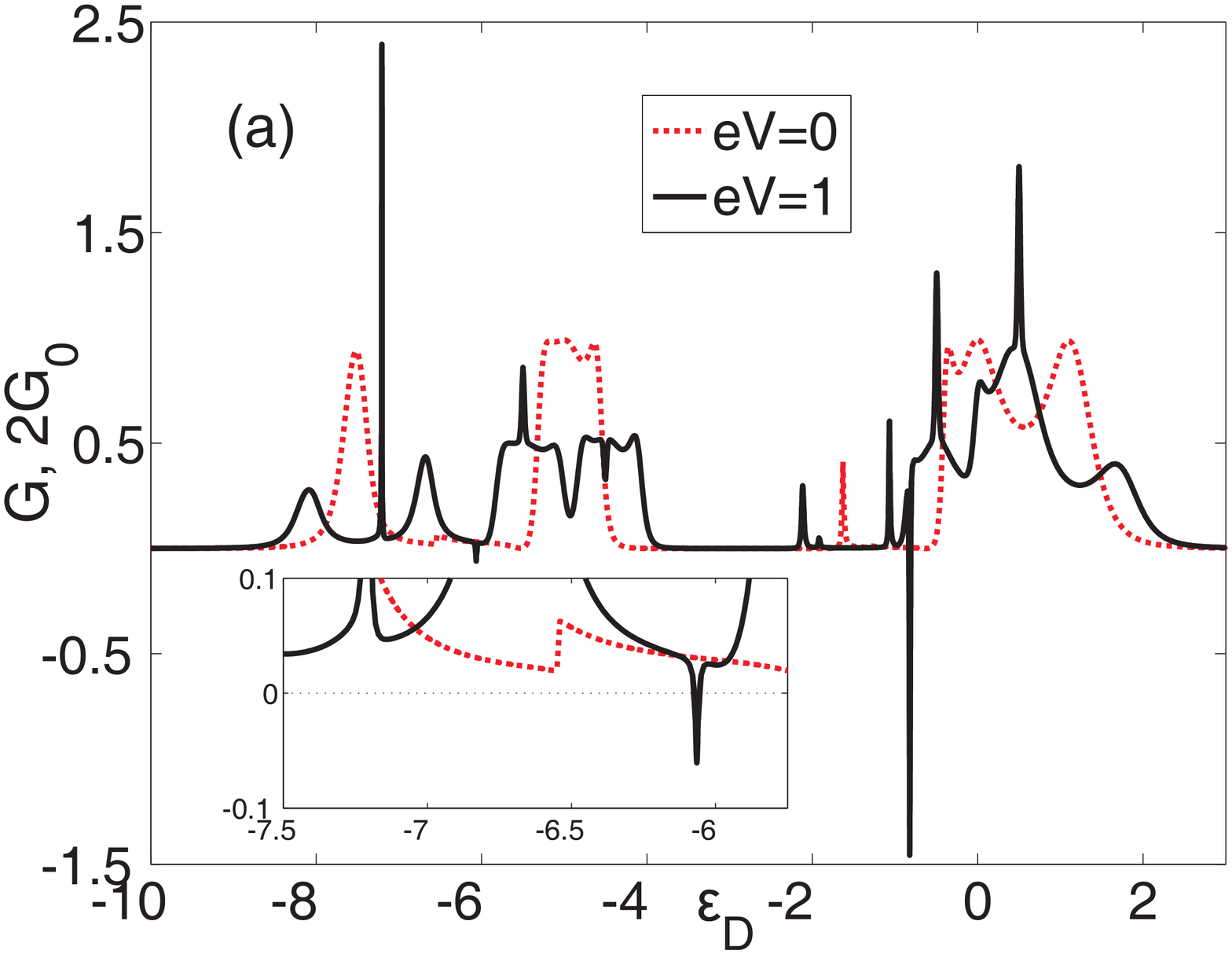}
\includegraphics[width=0.45\textwidth]{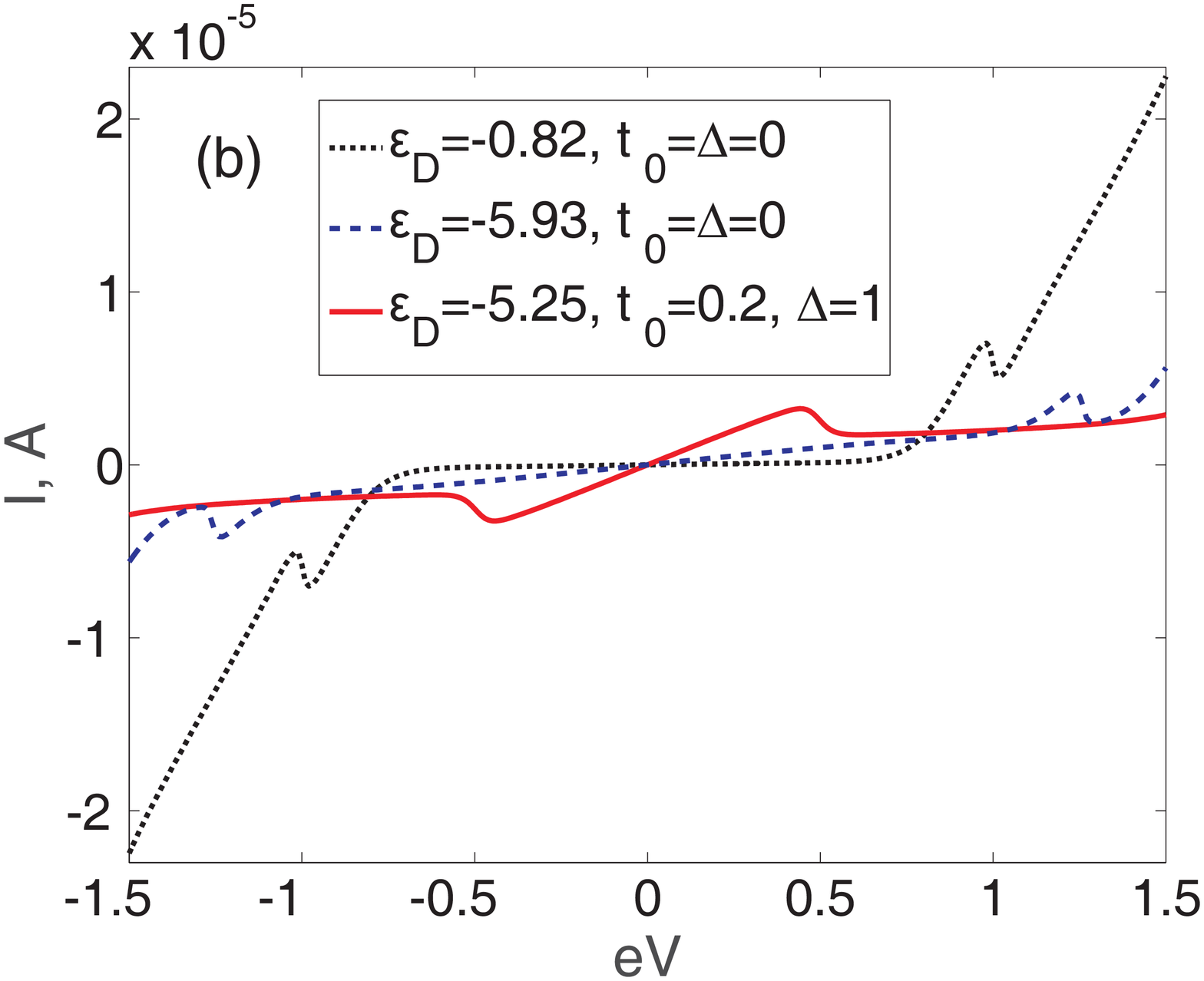}
\caption{The transport properties of anisotropic QQD. (a) The gate-voltage dependence of conductance. Inset: the Fano antiresonance and its splitting at $eV\neq0$. (b) The current-voltage characteristics. Parameters: $t_1=1$, $t_2=0.1$, the other parameters coincide with the ones used in the Fig. \ref{fig.2}.}\label{fig.3}
\end{figure}

Let us pass to the anisotropic situation, $t_1 \gg t_2$. The Figure \ref{fig.3}a represents the modification of gate-voltage dependence of conductance in this regime when the bias voltage is turned on. It is seen that the anisotropy of the kinetic processes in the QQD causes the appearance of conductance antiresonances with negative values. In the Figure \ref{fig.3}b the dotted curve shows the I-V characteristic in the gate field $\varepsilon_D=-0.82$ corresponding to the antiresonance of the highest amplitude in the Figure \ref{fig.3}a. The I-V curve has four sections where the behavior of conductance differs substantially. At source-drain field energies $\mid eV \mid \leq 0.75$, the current practically does not increase analogously to the Coulomb blockade effect. At $0.75 \leq \mid eV \mid \leq 1$ the significant growth takes place followed by a sharp decline at $\mid eV \mid \approx 1$ with a narrow valley. At $1 \leq \mid eV \mid \leq 1.5$ the current considerably increases as well as in the second section. The peak-to-valley ratio in this case is $\sim 1.4$. The similar scenario is observed if the QQD occupation is above half-filling (dashed curve in Fig. \ref{fig.3}b). The peak-to-valley ratio can be additionally increased if we take into account the hopping between the central dots and make their single-electron energies different by means of several gate electrodes ($t_0 \neq0$, $\Delta\neq0$). The I-V characteristic corresponding to this case is represented by a solid curve in the Figure \ref{fig.3}b. It is clearly seen that the valley is wider and the peak-to-valley ratio is $\sim 1.9$. In the situation of T-shaped QQD geometry ($t_2=0$) the peak-to-valley ratio is $\sim 2.6$. For the $\Gamma \ll U,~V$ mode and using the same relations between the hopping parameters $t_1,~t_2,~t_0$ as in the Figure \ref{fig.3} we can get the ratio of the order of $4$ (the last two cases are not represented in the Figure \ref{fig.3}).

\begin{figure}\center
\includegraphics[width=0.45\textwidth]{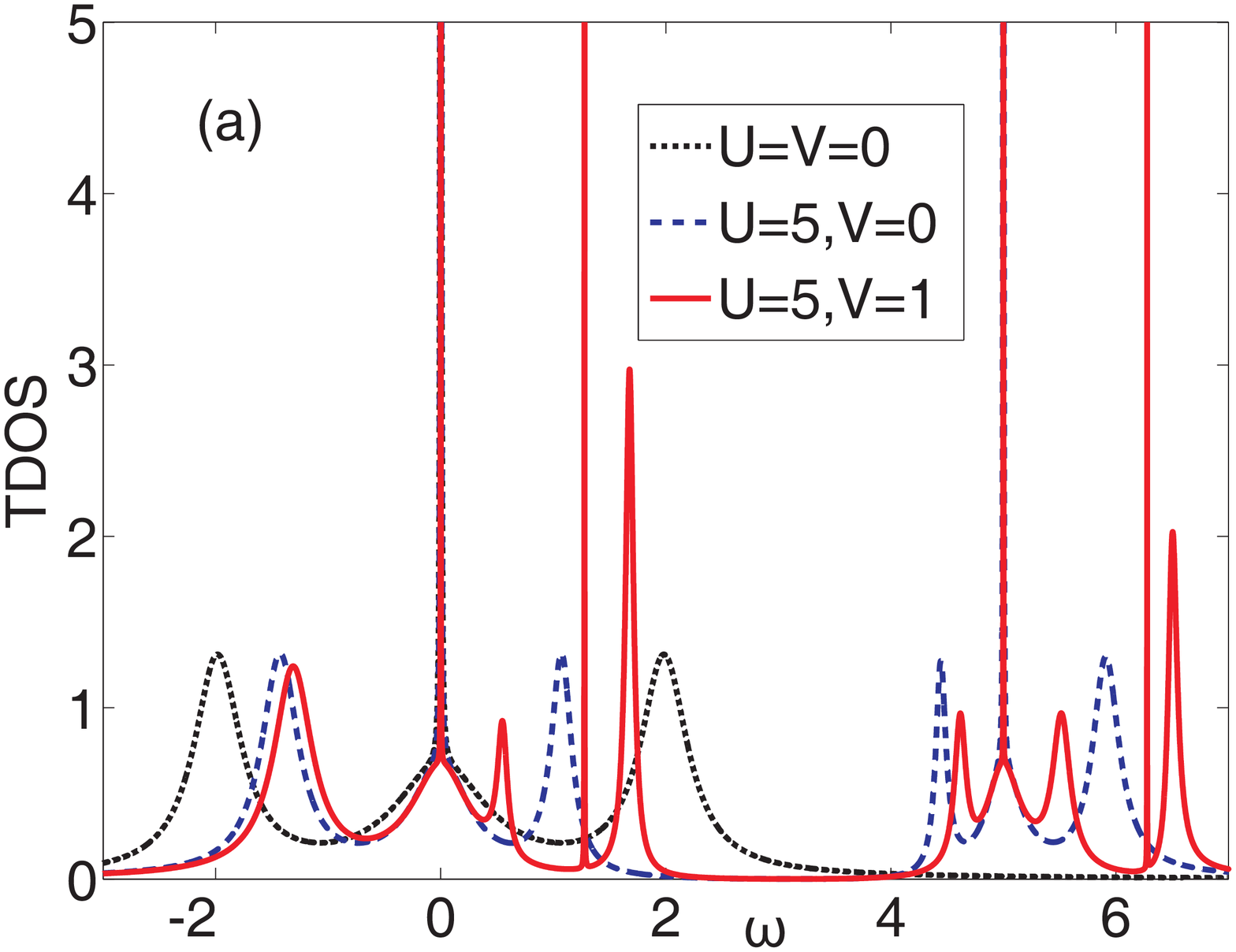}
\includegraphics[width=0.45\textwidth]{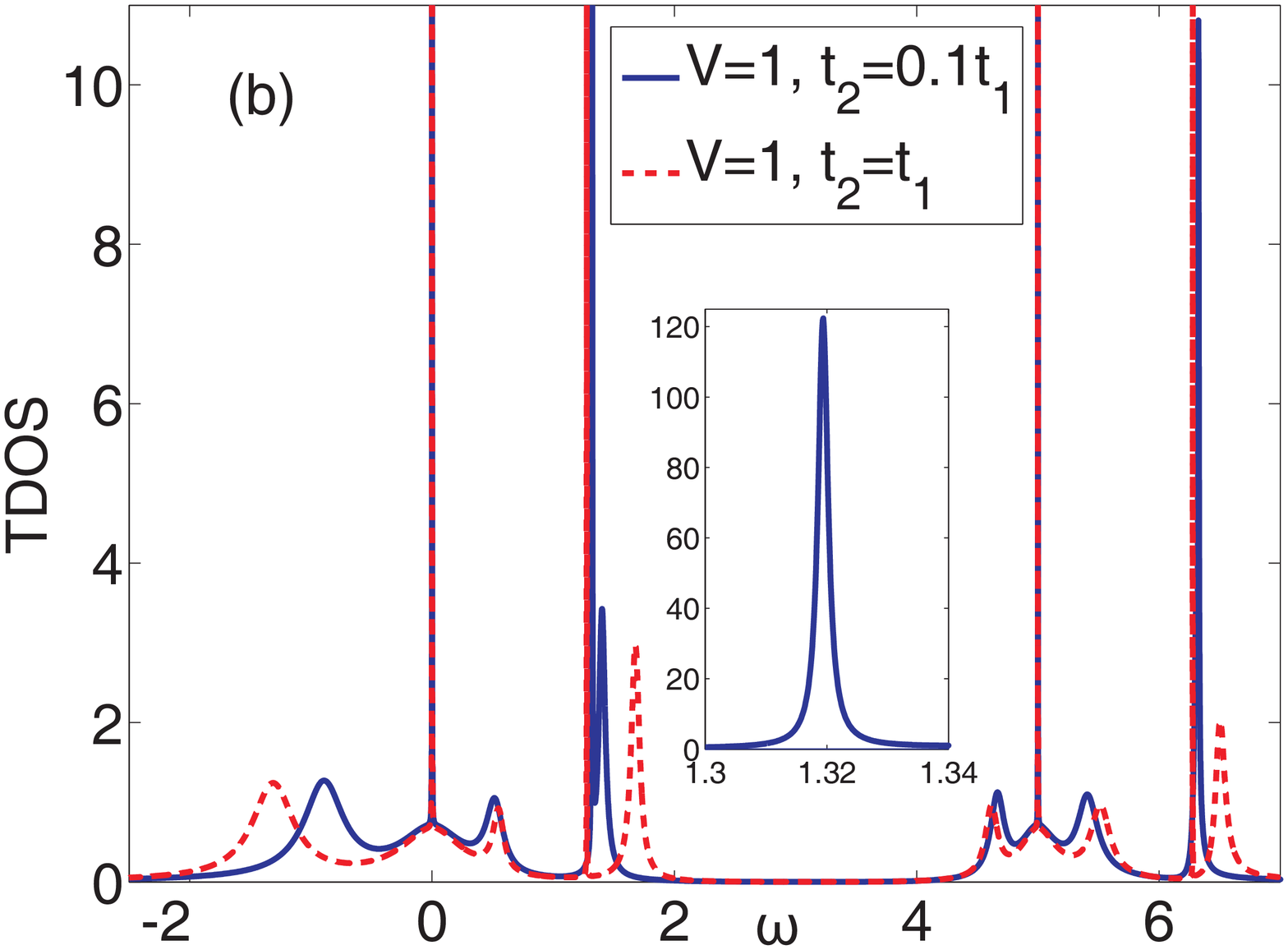}
\caption{(a) The influence of Coulomb correlations on the BICs in the density of states of isotropic QQD. (b) The effect of anisotropy of kinetic processes in the QQD on the BICs. Inset: one of the maxima related to the BIC which is formed at $V\neq0$. Parameters: $\varepsilon_D=0$, the other parameters coincide with the ones used in Fig. \ref{fig.3}.}\label{fig.4}
\end{figure}

The observed NDC effect is related to the features of density of states (DOS) of the QQD in the anisotropic regime, $TDOS\left(\omega\right)=-\frac{1}{\pi}\sum\limits_{i=1}^{4}Im\{G^{r}_{ii}\left(\omega\right)\}$. First, we start with the isotropic situation. The appropriate DOS is displayed in the Figure \ref{fig.4}a. In the absence of Coulomb interactions the positions of maxima of $TDOS\left(\omega\right)$ are determined by the energies of eigenstates of Hamiltonian $H_{QQD}\left(U=V=0\right)$ (dotted curve in Fig. \ref{fig.4}a). If $t_0=\Delta=0$ that there are four levels with the energies: $\varepsilon_D$, $\varepsilon_D\pm2t_1$. As it was shown in \cite{volya-03,sadreev-06}, the presence of the degeneracy can give rise to BICs. In our case the BIC is displayed by the infinitely narrow peak at $\omega=0$ whose width is characterized by the term $i\delta$ in $G^r_{ij}\left(\omega\right)$. Switching on the intradot Coulomb interactions results in the appearance of three new maxima due to the splitting of single-electron excitation energies of the individual dot: $\varepsilon_D$, $\varepsilon_D+U$ (dashed curve in Fig. \ref{fig.4}a). As a consequence, the additional BIC occurs \cite{sadreev-08}. The interdot Coulomb interaction causes the extra splitting of one-electron excitation energies. Thus, two new maxima and two BICs arise in the DOS (solid curve in Fig. \ref{fig.4}a). It is worth to note that these maxima are the reason of the conductance resonances in the linear response regime (dotted curve in Fig. \ref{fig.2}a). In particular, the induction of asymmetric Fano peaks at $V\neq0$ is attributed to the appearance of corresponding maxima in the dependence $TDOS\left(\omega\right)$ \cite{kagan-17a,kagan-17b}. In turn, the BICs do not manifest themselves in the QQD transport characteristics.

In the anisotropic situation the lifetime of two BICs induced by the interdot Coulomb correlations becomes finite. As a result, two narrow peaks of finite height emerge (solid curve in Fig. \ref{fig.4}b and the inset) and new Fano antiresonances appears in the conductance. One of them is shown in the inset of Figure \ref{fig.3}a at $\varepsilon_D\approx-6.5$ (see dotted curve). The nonzero value of $G$ is due to the temperature factor. It was already mentioned above that the conductance resonances are split in the nonequilibrium regime. In turn, the antiresonance under consideration is transformed into a narrow resonance and antiresonance with $G>0$ and $G<0$, respectively. They are placed at the distance of approximately $eV$ (in the inset of Fig. \ref{fig.3}a the bottom of resonance and the antiresonance at $eV\neq0$ are plotted by solid curve). The increase of bias voltage shifts the antiresonance to the right. Simultaneously, the Fano asymmetric peak arising at $V\neq0$ in the isotropic case is shifted to the left. Thus, the amplification of NDC is observed when preformed Fano features are close to each other and interact. The described scenario is also realized if the QQD occupation is less than one half.

Note that in \cite{arseev-12}, where a parallel-coupled double quantum dot is studied, the NDC effect induced by the Coulomb correlations occurs if the dots are connected with the leads asymmetrically. In our case the NDC takes place in the symmetric coupling regime. At the same time, the asymmetry of kinetic processes, leading to the above-mentioned peculiarities in the DOS and specific redistribution of dots' occupations, is a property of the device itself.

\section{Conclusion}

In this article we investigated the influence of nonequilibrium effects on quantum transport in a system of four quantum dots taking into account the Coulomb correlations. To find the expression that describes the electron current the nonequilibrium Green's functions and equation-of-motion technique are applied. In the last case the equations for the third-order Green’s functions were decoupled \cite{hubbard-64} as it had been described earlier in Refs. \cite{you-99a,kagan-17a,kagan-17b}. The numerical analysis of the QQD DOS showed that the system contains the BICs induced by the Coulomb interactions. It is shown that the anisotropy of kinetic processes in the QQD results in the finite lifetime of BICs which are created by the interdot Coulomb interaction. The consequent Fano antiresonances in the gate-voltage dependence of the conductance are shifted in nonequilibrium regime. The interaction of these features with the other Fano asymmetric peaks (which are caused by the interdot Coulomb correlations and appear even in the isotropic case) gives rise to significant enhancement of the NDC effect. It is demonstrated that the corresponding peak-to-valley ratio of the I-V characteristic can be significantly increased by the change of the system parameters.

We acknowledge fruitful discussions with P.I. Arseyev, N.S. Maslova and V.N. Mantsevich. This work was financially supported by the RFBR, Government of Krasnoyarsk Territory, Krasnoyarsk Region Science and Technology Support Fund, Projects No. 16-02-00073, No. 17-42-240441, and No. 17-02-00135. S.A. is grateful to the Council of the President of the Russian Federation for Support of Young Scientists and Leading Scientific Schools, project no. MK-3722.2018.2. The publication was prepared by M.Yu.K. within the framework of the Academic Fund Program at the National Research University Higher School of Economy in 2017-2018 (grant No. 18-05-0024) and by the Russian Academic Excellence Project "5-100".

\end{document}